%% file: conference_101719.tex
\newif\ifACMformat
\ACMformattrue
\PassOptionsToPackage{table}{xcolor}

\ifACMformat
  \documentclass[sigconf,screen=true,natbib=true,nonacm]{acmart}
  \input{acm-setting}

\else
  \documentclass[10pt, letter, conference]{IEEEtran}
  \IEEEoverridecommandlockouts
  \input{setting}

\fi
\ifACMformat
\else
  \usepackage{cite}
\fi

\newcommand{\legend}[0]{{\texttt{LeGend}}\xspace}
\def\BibTeX{{\rm B\kern-.05em{\sc i\kern-.025em b}\kern-.08em
    T\kern-.1667em\lower.7ex\hbox{E}\kern-.125emX}}
\newlist{myitemize}{itemize}{1}
\setlist[myitemize]{
  label=\textbullet, 
}

\usepackage{etoolbox}
\usepackage{dsfont}

\makeatletter
\patchcmd{\@makecaption}
  {\scshape}
  {}
  {}
  {}
\makeatother

\def\BibTeX{{\rm B\kern-.05em{\sc i\kern-.025em b}\kern-.08em
    T\kern-.1667em\lower.7ex\hbox{E}\kern-.125emX}}

\begin{document}

\title{\legend: A Data-Driven Framework for \underline{Le}mma \underline{Gen}eration in Har\underline{d}ware Model Checking
}




%

\ifACMformat
\author{Mingkai Miao}
\authornote{These authors contributed equally to this work.}
\affiliation{%
  \institution{HKUST(GZ)}
}
\email{mmiao815@connect.hkust-gz.edu.cn}

\author{Guangyu Hu}
\authornotemark[1]
\affiliation{%
  \institution{HKUST}
}
\email{ghuae@connect.ust.hk}

\author{Wei Zhang}
\affiliation{%
  \institution{HKUST}
}
\email{wei.zhang@ust.hk}

\author{Hongce Zhang}
\affiliation{%
  \institution{HKUST(GZ)}
}
\email{hongcezh@hkust-gz.edu.cn}
\else
\author{%
\IEEEauthorblockN{Mingkai Miao\textsuperscript{*}, Guangyu Hu\textsuperscript{*}, Wei Zhang, Hongce Zhang}
\IEEEauthorblockA{\textit{HKUST(GZ)}\\
Mingkai Miao: mmiao815@connect.hkust-gz.edu.cn\\
Hongce Zhang: hongcezh@hkust-gz.edu.cn}
\IEEEauthorblockA{\textit{HKUST}\\
Guangyu Hu: ghuae@connect.ust.hk\\
Wei Zhang: wei.zhang@ust.hk}
\thanks{\textsuperscript{*}These authors contributed equally to this work.}
}
\fi
\input{sections/abstract}

\maketitle

\input{sections/intro.tex}

\input{sections/preliminary}

\input{sections/method}
\input{sections/experiment}
\input{sections/conclusion}
\newpage
\bibliography{refs}
\balance

\end{document}

%% file: acm-setting.tex
\usepackage{amsmath,amsfonts}
\usepackage{adjustbox}
\usepackage{colortbl}
\usepackage{array} 
\usepackage{multirow}
\usepackage{multicol}
\usepackage{amsthm}
\usepackage{listings}
\usepackage[linesnumbered,ruled,vlined]{algorithm2e}
\usepackage{graphicx}
\usepackage{pgfplots}
\usepackage{textcomp}
\usepackage[table]{xcolor}
\definecolor{deepred}{RGB}{163, 0, 0}   
\definecolor{goodgreen}{RGB}{0, 120, 0}
\newcommand{\cmark}{\ding{51}}%
\newcommand{\xmark}{\ding{55}}%
\usepackage{cleveref}
\usepackage{makecell}
\usepackage{booktabs}
\usepackage{comment}
\usepackage{tabularx}
\usepackage{multirow}
\usepackage{bigdelim}
\usepackage{bm}
\usepackage{url}
\usepackage{xspace}
\usepackage{pifont}
\usepackage{arydshln}
\usepackage{verbatim}
\usepackage[referable]{threeparttablex}
\usepackage{calc} 
\usepackage{extarrows}
\usepackage{color}
\usepackage{float}
\usepackage{subcaption}
\usepackage{nicematrix}
\usepackage{geometry}
\usepackage[table]{xcolor}
\DeclareUnicodeCharacter{2011}{-}
\usepackage[utf8]{inputenc}
\usepackage{hhline}
\usepackage{booktabs} 
\usepackage{tabularx} 
\usepackage{tikz}
\usetikzlibrary{shapes.geometric, arrows}
\tikzstyle{startstop} = [rectangle, rounded corners, minimum width=3cm, minimum height=1cm,text centered, draw=black]
\tikzstyle{io} = [trapezium, trapezium left angle=70, trapezium right angle=110, minimum width=3cm, minimum height=1cm, text centered, draw=black]
\tikzstyle{process} = [rectangle, minimum width=3cm, minimum height=1cm, text centered, draw=black]
\tikzstyle{arrow} = [thick,->,>=stealth]

\usepackage{filecontents}                                  
\usepackage{pgfplots}
\usepackage{pgfplotstable}
\usepackage{scalefnt}

\definecolor{USTgold}{RGB}{153,102,0}
\definecolor{USTyellow}{RGB}{204,153,0}
\definecolor{USTyellowlight}{RGB}{255,212,0}
\definecolor{USTorange}{RGB}{255,166,26}
\definecolor{USTpink}{RGB}{255,157,157}
\definecolor{USTblue}{RGB}{0,51,102}
\definecolor{USTmiddle}{RGB}{0,116,188}
\definecolor{USTlight}{RGB}{99,202,225}
\definecolor{USTgray}{RGB}{204,204,204}
\definecolor{USTred}{RGB}{237,27,47}
\definecolor{USTdarkred}{RGB}{124,35,72}

\definecolor{CUHKorange}{RGB}{244,106,18} 
\definecolor{CUHKblue}{RGB}{0,111,190}    
\definecolor{CUHKgreen}{RGB}{0,127,128}   
\definecolor{CUHKred}{RGB}{228,46,36}     
\definecolor{CUHKyellow}{RGB}{198,148,34} 
\definecolor{CUHKdark}{RGB}{114,44,114}   
\definecolor{CUHKmiddle}{RGB}{144,44,144} 
\definecolor{CUHKlight}{RGB}{167,44,167} 
\definecolor{lightblue}{RGB}{223, 235, 247}

\newtheorem{theorem}{Theorem}
\newtheorem{corollary}{Corollary}
\newtheorem{lemma}[theorem]{Lemma}

\iftrue
\def\BibTeX{{\rm B\kern-.05em{\sc i\kern-.025em b}\kern-.08em
    T\kern-.1667em\lower.7ex\hbox{E}\kern-.125emX}}

\fi

\iftrue
\fi

\iftrue
\usepackage{titlesec}
\titlespacing\section{2pt}{5pt plus 1pt minus 1pt}{0pt plus 1pt minus 1pt}
\titlespacing\subsection{2pt}{5pt plus 1pt minus 1pt}{0pt plus 1pt minus 1pt}
\titlespacing\subsubsection{2pt}{5pt plus 1pt minus 1pt}{2pt plus 1pt minus 1pt}
\usepackage[inline]{enumitem}
\setlist{leftmargin=5.08mm}
\fi

\iftrue
\fi

%% file: setting.tex
\usepackage{amsmath,amsfonts}
\usepackage{colortbl}
\usepackage{array} 
\usepackage{multirow}
\usepackage{multicol}
\usepackage{amsthm}
\usepackage{listings}
\usepackage[linesnumbered,ruled,vlined]{algorithm2e}
\usepackage{graphicx}
\usepackage{pgfplots}
\usepackage{textcomp}
\usepackage{xcolor}
\usepackage{cleveref}
\usepackage{makecell}
\usepackage{booktabs}
\usepackage{comment}
\usepackage{tabularx}
\usepackage{adjustbox}
\usepackage{multirow}
\usepackage{bigdelim}
\usepackage{bm}
\usepackage{url}
\usepackage{xspace}
\usepackage{pifont}
\usepackage{arydshln}
\usepackage{verbatim}
\usepackage[referable]{threeparttablex}
\usepackage{calc} 
\usepackage{extarrows}
\usepackage{color}
\usepackage{float}
\usepackage{subcaption}
\usepackage{nicematrix}
\usepackage{geometry}
\usepackage{balance}
\usepackage{dsfont}
\usepackage{pifont}
\newcommand{\cmark}{\ding{51}}%
\newcommand{\xmark}{\ding{55}}%
\usepackage{csquotes}
\MakeOuterQuote{"}
\usepackage[table]{xcolor}
\definecolor{deepred}{RGB}{163, 0, 0}   
\definecolor{goodgreen}{RGB}{0, 120, 0}
\usepackage[utf8]{inputenc}
\usepackage{hhline}
\usepackage{booktabs} 
\usepackage{tabularx} 
\usepackage{tikz}
\usetikzlibrary{shapes.geometric, arrows}
\tikzstyle{startstop} = [rectangle, rounded corners, minimum width=3cm, minimum height=1cm,text centered, draw=black]
\tikzstyle{io} = [trapezium, trapezium left angle=70, trapezium right angle=110, minimum width=3cm, minimum height=1cm, text centered, draw=black]
\tikzstyle{process} = [rectangle, minimum width=3cm, minimum height=1cm, text centered, draw=black]
\tikzstyle{arrow} = [thick,->,>=stealth]

\usepackage{filecontents}                                  
\usepackage{pgfplots}
\usepackage{pgfplotstable}
\usepackage{scalefnt}

\definecolor{USTgold}{RGB}{153,102,0}
\definecolor{USTyellow}{RGB}{204,153,0}
\definecolor{USTyellowlight}{RGB}{255,212,0}
\definecolor{USTorange}{RGB}{255,166,26}
\definecolor{USTpink}{RGB}{255,157,157}
\definecolor{USTblue}{RGB}{0,51,102}
\definecolor{USTmiddle}{RGB}{0,116,188}
\definecolor{USTlight}{RGB}{99,202,225}
\definecolor{USTgray}{RGB}{204,204,204}
\definecolor{USTred}{RGB}{237,27,47}
\definecolor{USTdarkred}{RGB}{124,35,72}

\definecolor{CUHKorange}{RGB}{244,106,18} 
\definecolor{CUHKblue}{RGB}{0,111,190}    
\definecolor{CUHKgreen}{RGB}{0,127,128}   
\definecolor{CUHKred}{RGB}{228,46,36}     
\definecolor{CUHKyellow}{RGB}{198,148,34} 
\definecolor{CUHKdark}{RGB}{114,44,114}   
\definecolor{CUHKmiddle}{RGB}{144,44,144} 
\definecolor{CUHKlight}{RGB}{167,44,167} 
\definecolor{lightblue}{RGB}{223, 235, 247}

\iftrue
\def\BibTeX{{\rm B\kern-.05em{\sc i\kern-.025em b}\kern-.08em
    T\kern-.1667em\lower.7ex\hbox{E}\kern-.125emX}}

\fi

\iftrue
\fi

\iftrue
\usepackage{titlesec}
\titlespacing\section{2pt}{5pt plus 1pt minus 1pt}{0pt plus 1pt minus 1pt}
\titlespacing\subsection{2pt}{5pt plus 1pt minus 1pt}{0pt plus 1pt minus 1pt}
\titlespacing\subsubsection{2pt}{5pt plus 1pt minus 1pt}{2pt plus 1pt minus 1pt}
\usepackage[inline]{enumitem}
\setlist{leftmargin=5.08mm}
\fi

\iftrue
\fi

%% file: sections/abstract.tex
\begin{abstract}

Property checking of RTL designs is a central task in formal verification. Among available engines, IC3/PDR is a widely used backbone whose performance critically depends on inductive generalization—the step that generalizes a concrete counterexample-to-induction (CTI) cube into a lemma. Prior work has explored machine learning to guide this step and achieved encouraging results, yet most methods adopt a per-clause graph analysis paradigm: for each clause they repeatedly build and analyze graphs, incurring heavy overhead and creating a scalability bottleneck. We introduce \legend, which replaces this paradigm with one-time global representation learning. \legend pre-trains a domain-adapted self-supervised model to produce latch embeddings that capture global circuit properties. These precomputed embeddings allow a lightweight model to predict high-quality lemmas with negligible overhead, effectively decoupling expensive learning from fast inference. Experiments show \legend accelerates two state-of-the-art IC3/PDR engines across a diverse set of benchmarks, presenting a promising path to scale up formal verification.\looseness=-1

\end{abstract}

%% file: sections/intro.tex
\section{Introduction}
\label{sec:intro}

Formal property verification is an indispensable pillar of modern integrated circuit design, with the Property-Directed Reachability (PDR) algorithm~\cite{pdr} (a.k.a. the IC3 algorithm~\cite{ic3}) standing as one of the commonly-used state-of-the-art model checking methods. The performance of PDR is critically dependent on \textit{inductive generalization}, the process of abstracting a specific counterexample into a general lemma (a Boolean clause). The quality of these lemmas directly dictates the algorithm's convergence speed, making it a prime target for optimization.

However, conventional inductive generalization methods are fundamentally limited, because they typically rely on localized greedy heuristics~\cite{Bradley07, Hassan13} that lack a global perspective of the design's behavior and often produce ``myopic'' lemmas that fail to generalize effectively~\cite{Hu21, krishnan2020global}. Such sub-optimal clauses often result in wasted computational effort, as the PDR algorithm may unnecessarily explore irrelevant parts of the state space led by the low-quality lemmas,
thus hindering the convergence speed to a formal proof. The central challenge, therefore, is how to develop a more globally-aware strategy for inductive generalization.\looseness=-1


To this end, researchers have turned to machine learning (ML)---specifically graph neural networks (GNNs)---to generate higher-quality lemmas. However, while pioneering works~\cite{Hu23,Hu21} in this domain demonstrate the potential of graph learning, they are built upon the same computationally expensive paradigm that requires \textbf{repetitive, instance-specific graph processing}. Specifically, NeuroPDR~\cite{Hu23} needs to construct and analyze a new graph for \textit{each} counterexample that the algorithm finds.  While DeepIC3~\cite{Hu21} decouples generation of candidate lemmas from the PDR algorithm itself, it still needs the repeated graph analysis for each of the thousands of \textit{candidate clauses} it evaluates offline.
On the other hand, IC3-CTP~\cite{predictinglemmasgeneralizationic3} relies on a lightweight, rule-based SAT-guided heuristic for improved lemmas. Although fast, it does not take good advantage of the features of circuits and would demand extensive solver-specific rewrites to port to a new  PDR implementation.
As summarized in \textbf{\Cref{tab:literature_comparison}}, none of these methods marries learning power with both low overhead and easy portability.

\input{tables/lemma_predictor_comparison}


This paper presents \legend, a framework that architects a new paradigm:
 \textit{one-time circuit embedding} for \textit{massively parallel offline lemma prediction}. 
Instead of analyzing countless individual clauses, \legend 
shifts most of the machine learning computations to
a single, upfront, and fully offline stage. 
It begins by constructing a global graph of the entire circuit. This step runs only once. 
It then computes a rich and dense vector embedding for every latch\footnote{In the literature of hardware model checking, state-holding elements are uniformly referred to as latches regardless of their circuit implementations.}, using a pre-trained self-supervised contrastive learning model. This embedding captures the global structural role and local dynamic properties (such as signal flip rates) of latches.




With these 
pre-computed embeddings, the bottleneck of prior methods is eliminated, because \legend no longer needs to build a graph for each of the thousands of sampled candidate clauses. Instead, a lightweight permutation-invariant model predicts the quality of a clause simply by operating on the embedding vectors of its constituent latches.
This process is orders of magnitude faster than the prior per-clause graph analysis. The final set of high-quality lemmas is then 
\textit{side-loaded}~\cite{Hu21} into the PDR solver at initialization, providing a powerful head start to the verification process. This runtime layer is solver-agnostic, allowing \legend\ to plug into any modern IC3/PDR engine with minimal code changes.


Our main contributions are as follows:
\begin{itemize}[leftmargin=*,noitemsep,topsep=0pt]
    \item This paper is this first work that addresses the unsustainable cost of per-clause graph processing in prior ML-guided PDR methods.
    \item We architect a new framework, \legend{}, centered on \textit{one-time global embedding}, decoupling expensive, one-off representation learning from fast, massively parallel inference.
    \item We devise a circuit‑aware contrastive pre-training scheme together with a permutation‑invariant latch‑set predictor, yielding high‑quality inductive lemmas while eliminating per‑clause graph processing.\looseness=-1
    \item We demonstrate 
    on the public hardware model checking competition (HWMCC) benchmarks that our approach provides a practical and scalable solution, significantly accelerating the state-of-the-art PDR implementations.
\end{itemize}

The remainder of this paper is organized as follows. Section~\ref{sec:preliminary} provides background on the PDR algorithm. Section~\ref{sec:method} details the \legend framework, including representation learning and offline prediction. Section~\ref{sec:experiment} presents our experimental setup and results. Finally, Section~\ref{sec:conclusion} concludes the paper and discusses future directions.

%% file: tables/lemma_predictor_comparison.tex
\begin{table}[tbp]
  \centering
  \caption{Key characteristics of modern representative lemma-prediction techniques for the IC3/PDR algorithm.}
  \label{tab:literature_comparison}
  \renewcommand{\arraystretch}{1.2}
  \begin{adjustbox}{width=\columnwidth,center}
    \begin{tabular}{lcccc}
      \toprule
      \textbf{Method} & \textbf{ML-aided} & \textbf{Circuit-aware} & \textbf{Portability} & \textbf{Overhead}\\
      \midrule
      NeuroPDR~\cite{Hu23} & \cmark & \cmark & \xmark & \textcolor{deepred}{Heavy}\\
      DeepIC3~\cite{Hu21}  & \cmark & \cmark & \cmark  & \textcolor{deepred}{Heavy}\\
      IC3-CTP~\cite{predictinglemmasgeneralizationic3}  & \xmark  & \xmark         & \xmark & \textcolor{goodgreen}{Light}\\
      \midrule
      \rowcolor{gray!20}
      \textbf{\legend} & \textbf{\cmark} & \textbf{\cmark} & \textbf{\cmark} & \textbf{\textcolor{goodgreen}{Light}}\\
      \bottomrule
    \end{tabular}
  \end{adjustbox}
\end{table}

%% file: sections/preliminary.tex
\section{Preliminary}\label{sec:preliminary}

\newcommand{\Block}{\texttt{Block}}
\newcommand{\Propagate}{\texttt{Propagate}}
\newcommand{\Generalize}{\texttt{Generalize}}

\subsection{The IC3/PDR Algorithm}
IC3/PDR~\cite{ic3,pdr} is a state-of-the-art algorithm for hardware model checking. It aims to prove a safety property $P$ by finding a safe \textit{inductive invariant}. The core idea is to build a sequence of clause sets, called frames, $F_0, F_1, \dots, F_k$, which are over-approximations of the states reachable in at most $k$ steps. $F_0$ is initialized with the initial states $I$, and for all $i \ge 0$, $F_i \Rightarrow P$ must hold. An inductive invariant is found if two consecutive frames become logically equivalent, i.e., $F_i = F_{i+1}$.
\input{algorithms/ic3_pdr_pseudocode}
The overall workflow of the algorithm is shown in \textbf{\Cref{alg:pdr}}. The main loop iteratively refines the frames. In the $k$-th iteration, it first checks if any state reachable in $k$ steps (approximated by $F_k$) can reach a bad state (a state violating $P$). If such a state 
is found, the algorithm calls the \Block~procedure (detailed in \textbf{\Cref{alg:block_and_generalize}}) to eliminate it. If blocking succeeds, the loop continues. Otherwise, it means a real counterexample trace has been found, and the property is deemed \texttt{Unsafe}. After ensuring no bad states are reachable from $F_k$, the algorithm attempts to propagate lemmas 
to later frames
(the \Propagate~function). If this results in a fixed-point ($F_i = F_{i+1}$), an inductive invariant has been discovered, and the property is \texttt{Safe}. 

\input{algorithms/inductive_generalization}

The core of IC3/PDR lies in the recursive blocking and generalization mechanism, shown in \textbf{\Cref{alg:block_and_generalize}}. When given a cube $c$ to block at frame $i$, the \Block~procedure checks if the corresponding lemma $\neg c$ is \textit{inductive relative} to the previous frame, $F_{i-1}$ (the function is defined in line~22-24). If not, it means there is a predecessor state $p$ (a Counterexample-To-Induction, or CTI~\cite{understanding_ic3}) in $F_{i-1}$ that can transition to $c$. The algorithm then makes a \textcolor{red}{recursive call to block this predecessor $p$ at frame $i-1$} (line 5 in \textbf{\Cref{alg:block_and_generalize}}). This recursive process continues until a blockable predecessor is found or the initial states (frame 0) are reached.\looseness=-1

Once a cube $c$ is proven to be inductive relative to a frame, the \Generalize~procedure is invoked~\cite{ic3}. As detailed in \textbf{\Cref{alg:block_and_generalize}}, this function attempts to simplify the cube by iteratively removing literals and checking if the resulted smaller cube is still (relatively) inductive. This process creates a more general lemma that covers a larger set of unreachable states. The final generalized lemma is then \textcolor{blue}{added to a range of frames} (line 11 in \textbf{\Cref{alg:block_and_generalize}}), strengthening the over-approximations. Literal removal in this step could follow different strategies, such as CTG~\cite{Hassan13} and EXCTG~\cite{exctg_su}. The quality of this generalization step is critical to the algorithm's performance.


\subsection{Clause side-loading}
\input{algorithms/side-loading_pseudocode}
\label{sec:side-loading}
To enhance the efficiency of the inductive proof search, IC3/PDR can be augmented with a technique known as \textit{clause side-loading}~\cite{Hu21} as formalized in \textbf{\Cref{alg:sideloading_full}}. This approach aims to accelerate convergence to a fixed-point by injecting a set of pre-computed lemmas into the frames at the beginning of the verification process. Namely, at initialization on line~6-7 in \textbf{\Cref{alg:sideloading_full}} (corresponding to line~1 in \textbf{\Cref{alg:pdr}}), $F_1$ could be set to the side-loaded clauses instead of an empty set (representing $\top$), as long as the clauses comply with the constraints over the frames~\cite{pdr}. These clauses will later be carried forward by the \Propagate function and may eventually become part of the inductive invariant.
While seemingly promising, 
side-loading may not always be beneficial: its effectiveness is critically contingent on the quality of the injected lemmas.

On one hand, high-quality inductive clauses can significantly prune the search space. Early work has demonstrated that even a small number of clauses sampled from a known inductive invariant can lead to substantial speedups~\cite{Hu21}. This suggests that if an oracle could provide perfectly generalized lemmas, the performance of PDR would be greatly enhanced.

However, on the other hand, side-loading a large number of irrelevant clauses without careful selection slows down the algorithm, because the solver will need to spend more time per SAT query and the IC3/PDR state‑space search will be steered toward unproductive areas~\cite{Hassan13}.
Consequently, the core challenge of side-loading is not merely to generate a large set of candidate clauses, but to meticulously select a smaller set of \textit{high-quality} clauses.
This central problem is the focus of \legend, which maps the vast pool of candidate 
lemmas
$S_{\text{candidates}}$ 
to a refined high-quality set $S_{\text{ext}}$ (line~3 in \textbf{\Cref{alg:sideloading_full}}). 
Then, the subsequent injection follows the standard side-loading interface used in DeepIC3~\cite{Hu21}. 

%% file: algorithms/ic3_pdr_pseudocode.tex
\begin{algorithm}[b!]
\small
  \caption{The IC3/PDR Algorithm Overview}
  \label{alg:pdr}
  \SetAlgoLined
  \SetKwFunction{Block}{BlockCube}
  \SetKwFunction{Propagate}{PropagateLemmas}
  \SetKwFunction{GetModel}{GetWitness}
  \SetKwInOut{Input}{Input}
  \SetKwInOut{Output}{Output}
  \BlankLine
  
  \Input{A safety problem $\langle I, T, P \rangle$, where $I$: initial states,  $T$: transition relation, $P$: the safety property to prove }
  \Output{\texttt{Safe} or \texttt{Unsafe}}
  \BlankLine

  $F_0 \gets I$; $F_1 \gets \top$ ; $k \gets 1$\; 
  \While{\texttt{true}}{
    \While{$F_k \land \neg P$ is SAT with witness $c$}{
        \If{\textbf{not} \Block{$c$, $k$}}{
            \KwRet{\texttt{Unsafe}}\; \tcp{Found a path to a bad state}
        }
    }
    \BlankLine
    \tcp{Push lemmas forward to find an invariant}
    \If{\Propagate{$k$}}{
        \KwRet{\texttt{Safe}}\; \tcp{Found an inductive invariant}
    }
    \BlankLine
    $k \gets k+1$; $F_k \gets \top$\; \tcp{Increment frame}
  }
  
  \BlankLine
  \hrule
  \BlankLine
  \SetKwProg{Fn}{Function}{:}{}
  \SetKwFunction{IsRelInductive}{IsRelativeInductive}
  \Fn{\Propagate{$k_{max}$}}{
      \For{$i = 1 \to k_{max}-1$}{
          \ForEach{lemma $\neg c \in F_i \setminus F_{i+1}$}{
              \If{\IsRelInductive{$c, i$}}{
                  $F_{i+1} \gets F_{i+1} \land \neg c$\;
              }
          }
      }
      \lIf{$\exists i < k_{max}$ s.t. $F_i = F_{i+1}$}{\Return \texttt{true}}
      \lElse{\Return \texttt{false}}
  }
\end{algorithm}

%% file: algorithms/inductive_generalization.tex
\begin{algorithm}[t!]
\small

    \caption{Recursive Blocking and Generalization}
    \label{alg:block_and_generalize}
    \SetAlgoLined
    \SetKwFunction{IsRelInductive}{IsRelativeInductive}
    \SetKwFunction{GetPred}{GetPredecessor}
    \SetKwFunction{Generalize}{Generalize}
    \SetKwFunction{Block}{BlockCube}
    \SetKwProg{Fn}{Function}{:}{}
    \BlankLine
    
    \Fn{\Block{$c$, $i$}}{
        \lIf{$i=0$}{\Return \texttt{false}}
        
        \While{\textbf{not} \IsRelInductive{$c, i-1$}}{
            $p \gets \GetPred()$\; \tcp{Get predecessor}
            \If{\textbf{not} \textcolor{red}{\Block{$p, i-1$}}}{
                \Return \texttt{false}\; \tcp{Predecessor cannot be blocked}
            }
        }
        
        \tcp{$\neg c$ is inductive relative to $F_{i-1}$}
        $g \gets \Generalize(c, i-1)$\;

        \For{$j=1 \to i$}{
            \textcolor{blue}{$F_j \gets F_j \land \neg g$}\; \tcp{Add generalized lemma}
        }
        \Return \texttt{true}\;
    }
    
    \BlankLine
    \hrule
    \BlankLine

    \Fn{\Generalize{$c$, $i$}}{
        \ForEach{literal $l \in c$}{
             $c_{cand} \gets c \setminus \{l\}$\;
             \tcp{Check if smaller cube is still inductive}
             \If{\IsRelInductive{$c_{cand}, i$}}{
                 $c \gets c_{cand}$\; \tcp{Drop literal}
             }
        }
        \Return{$c$}\; \tcp{Return minimized cube}
    }

    \BlankLine
    \hrule
    \BlankLine

    \Fn{\IsRelInductive{$c, i$}}{
        \tcp{Is $\neg c$ inductive relative to $F_i$?}
        \lIf{$F_i \land \neg c \land T \land c'$ is SAT}{\Return \texttt{false}}
        \lElse{\Return \texttt{true}}
    }
\end{algorithm}

%% file: algorithms/side-loading_pseudocode.tex
\begin{algorithm}[h]
\small
   
\caption{The Clause Sideloading Procedure}
\label{alg:sideloading_full}
\SetKwFunction{FGenLemmas}{GeneratePotentialLemmas}
\SetKwFunction{FBlock}{Block}
\SetKwFunction{FGetCube}{ModelToCube}
\SetKwFunction{FGeneralize}{GeneralizeCTI}
\SetKwFunction{FMainICThree}{SideloadingEnhancedPDR}
\SetKwFunction{FLegendSelect}{\legend}
\SetKwData{Frames}{F}

\SetKwBlock{PhaseOne}{Lemma Generation}{end}
\SetKwBlock{PhaseTwo}{Integration with PDR: \FMainICThree{$I, T, P$}}{end}

\PhaseTwo{
    $S_{candidates} \leftarrow \FGenLemmas(P, T)$\;
    \tcp*[h]{Our work: generate quality lemmas}\;
    \textcolor{blue}{$S_{ext} \leftarrow \FLegendSelect(S_{candidates})$}\; 
    $\Frames_0 \leftarrow I$\;
    \lIf{$(\text{sat}, s) \leftarrow \text{Solve}(\Frames_0 \land \neg P)$}{\KwRet \texttt{Unsafe}}
    $\Frames_1, \dots, \Frames_{k} \leftarrow \emptyset$\;
    \tcp*[h]{Sideload high-quality lemmas}\;
    $\Frames_1 \leftarrow \Frames_1 \cup S_{ext}$\; 
    
  \While{\texttt{true}}{
    \While{$F_k \land \neg P$ is SAT with witness $c$}{
        \lIf{\textbf{not} \Block{$c$, $k$}}{
            \KwRet{\texttt{Unsafe}} 
        }
    }
    ... \tcp*[h]{Main PDR loop continues}\;
    }
}
\end{algorithm}

%% file: sections/method.tex
\section{Our Method}
\label{sec:method}

\begin{figure*}[htbp]
    \centering
    \includegraphics[width=0.95\textwidth]{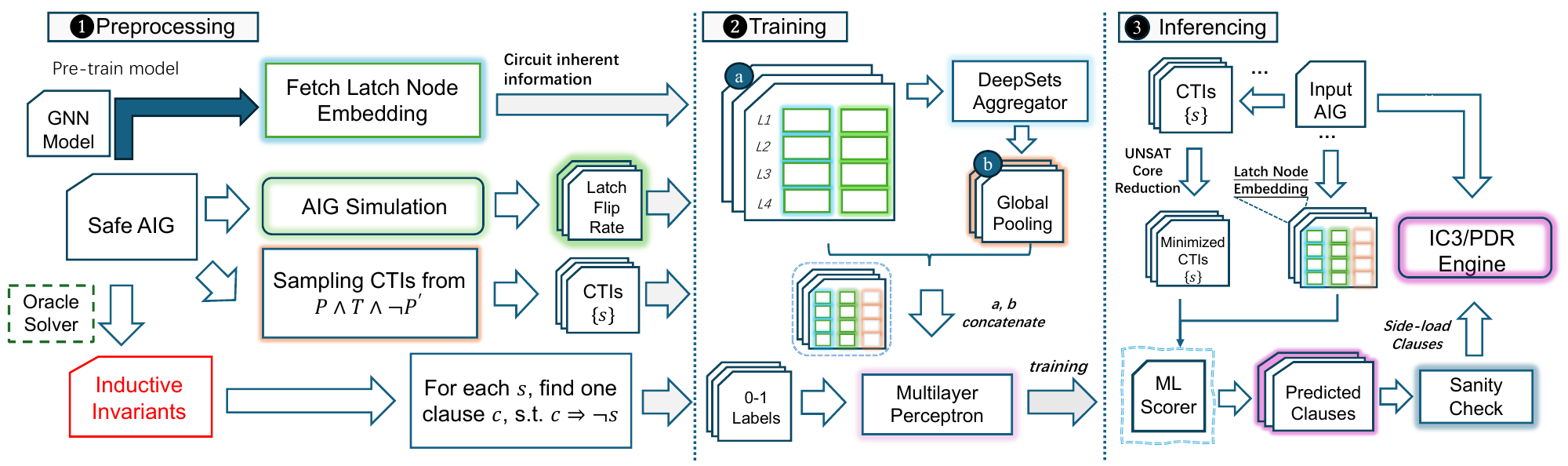}

    \caption{Overview of the \legend\ pipeline with three stages.
    \textbf{Stage 1: Preprocessing.} For each circuit we build a global graph and run a pre‑trained self‑supervised GNN once offline to obtain per‑latch embeddings, which are cached and augmented with a dynamic flip‑rate feature. An oracle solver finds the inductive invariants (as a set of clauses) for those instances in the training set.
    \textbf{Stage 2: Training.} A permutation‑invariant DeepSets aggregator with global pooling produces a clause‑context vector, and a lightweight MLP scores \emph{each literal} conditioned on this context.
    \textbf{Stage 3: Inference.} For an AIG file, the encoder is kept frozen and is executed once to obtain that circuit’s per‑latch embeddings; these precomputed vectors are then reused across all candidate clauses within the same circuit, so there is \emph{no per‑clause} graph construction, graph convolutions, or message passing. CTIs are sampled and minimally reduced, literal scores are thresholded to assemble predicted clauses, and clauses that pass the sanity checks (initiation and first-step consistency) are side‑loaded into the IC3/PDR solver.}
    \label{fig:overview}
    \vspace{-12pt}
\end{figure*}


As mentioned in Section~\ref{sec:intro}, prior ML‑guided PDR relies on per‑clause graph analysis: a fresh graph needs to be built for every candidate clause or counterexample~\cite{Hu21, Hu23}. 
This approach is not only computationally intense but also conceptually limited, as it only captures localized circuit properties, failing to leverage the global structural context of the design.

To break this barrier, we introduce \legend, a framework built on the paradigm of \textit{one-time global representation learning} for \textit{massively parallel lemma prediction}. As illustrated in \textbf{\Cref{fig:overview}}, \legend~architecturally decouples the expensive graph-based learning from the fast per-clause inference. This is achieved through a multi-stage pipeline designed to maximize both efficiency and quality of the generated lemmas.

\subsection{Preparation: CTI Generation and Minimization}

We start by generating a large initial pool of candidate clauses via sampling Counterexamples-To-Induction (CTIs)~\cite{ic3} from the formula $P\wedge T \wedge P'$.
Specifically, we develop a high-performance model sampling engine that leverages the SOTA SAT solver, enabling efficient generation of distinct CTIs.
The models are then reduced using UNSAT core minimization following the standard practice in the prior work~\cite{zhang2003extracting}.
The resulted minimized CTIs provide the starting point for later inductive generalization. This step is needed for both training and testing.

\subsection{Global Representation Learning}
The objective of representation learning is to produce a  vector representation (the \textit{embedding}) to represent the structural and functional information for each latch in the circuit. 
The encoder is trained separately on a corpus of circuits. At inference on a new circuit, we run a one‑time forward pass with this pre-trained encoder to obtain all latch embeddings. This per‑circuit embedding is the only graph‑based computation during solving. Subsequent literal and clause scoring uses these embeddings only without accessing the circuit graphs.\looseness=-1


\subsubsection{From AIGER Circuit to Graph Representation.}
Prior methods rely on constructing a small, localized graph for each candidate clause~\cite{Hu21}. This approach inherently limits the model's receptive field. A latch's importance is not just defined by its immediate neighbors but also by its position within the entire circuit. Therefore, our framework begins by constructing a single monolithic graph that represents the entire circuit topology from the gate-level AIGER netlist\footnote{We utilize the open-source \texttt{aigverse}~\cite{aigverse} library for parsing sequential AIGER files and converting them into a graph representation.}. This global view allows our machine learning model to capture long-range dependencies and high-order structural information that is fundamentally inaccessible to per-clause local analyses.

\subsubsection{Training a Latch Encoder.}
The ultimate goal of \legend is to predict potential clauses for IC3/PDR, which are unordered sets of literals. An effective prediction model requires an embedding for each latch that encodes not only its \textit{individual structural importance} but also its \textit{relational properties} when grouped with other latches in a clause. To this end, we design a self-supervised task to train a encoder model to generate these  embeddings without direct supervision, as ``good'' lemmas are not known \textit{a priori}.
\begin{figure}[b!]
    \centering
    \includegraphics[width=0.9\columnwidth]{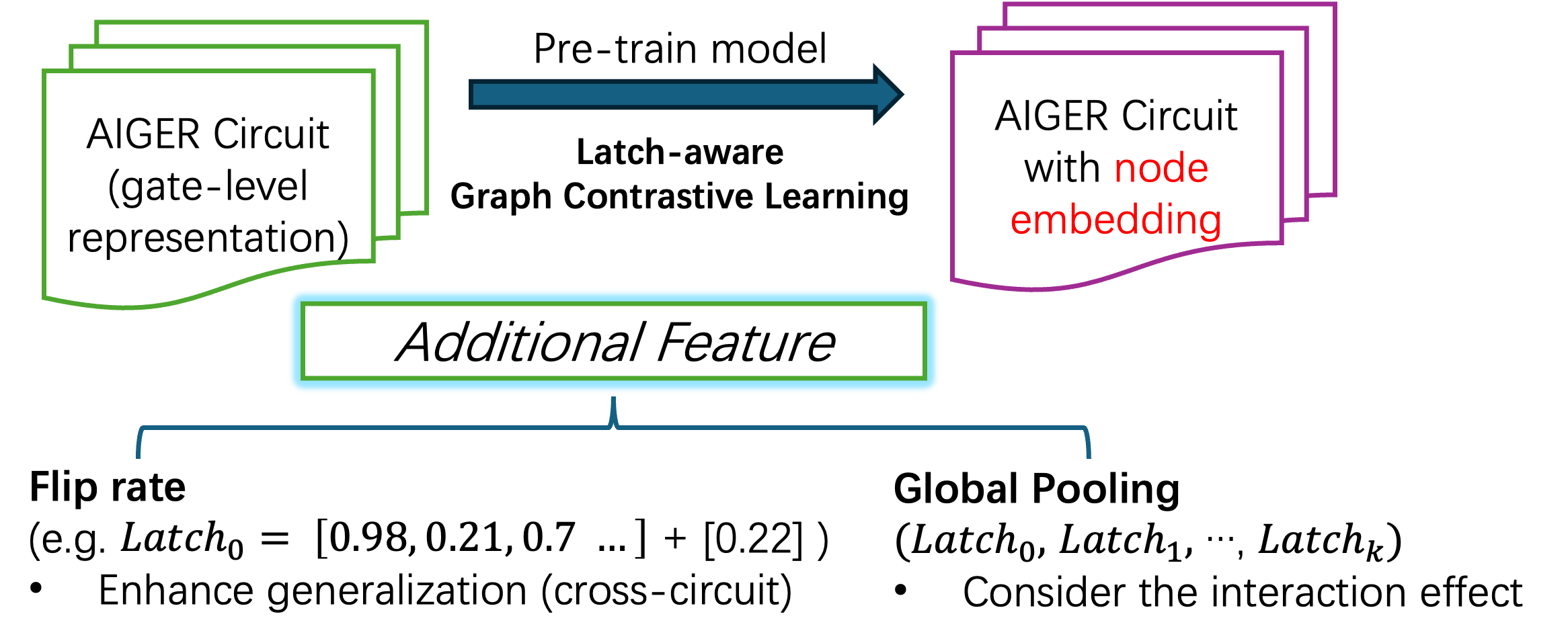}
    \caption{The latch embedding pipeline. We start with a gate-level AIGER circuit, which is converted into a graph, and we use a pre-trained GraphCL model to generate structural latch embeddings. These are further enhanced with signal flip rates, the dynamic functional feature to capture both static structure and dynamic behavior.}
    \label{fig:latch2embedding}
\end{figure}

\textbf{\Cref{fig:latch2embedding}} illustrates the self-supervised contrastive learning task that we design for the hardware model checking problems.
The core principle is that a latch's embedding should be invariant to minor perturbations that do not alter its fundamental function, yet sensitive enough to distinguish it from functionally distinct latches. We instantiate this principle by adapting the Graph Contrastive Learning (GraphCL) framework~\cite{graphcl} (self‑supervised pre-training that maximizes agreement between two perturbed views of the same circuit graph). Instead of using generic augmentations, we select augmentations that mimic plausible structural differences in a circuit.

Specifically, for each latch (referred to as the ``anchor''), we generate two augmented graphs (the ``views'') to form a \textit{positive pair}, while treating all other latches in a batch as \textit{negative pairs}. A Graph Isomorphism Network (GIN) encoder~\cite{gin} is then trained to maximize embedding similarity for positive pairs and minimize it for negative pairs, using the NT-Xent (the normalized
temperature-scaled cross entropy loss)~\cite{simclr}:
\begin{equation}
\mathcal{L}_{\text{NT-Xent}} = -\log\frac{\exp(\text{sim}(z_i, z_p) / \tau)}{\sum_{j=1}^{N} \exp(\text{sim}(z_i, z_j) / \tau)}
\end{equation}
where $z_i$ is the anchor embedding, $z_p$ is its positive pair, and $\tau$ is a temperature hyperparameter. Our contribution lies in the domain-specific adaptations of the data augmentations, including:
\begin{itemize}[leftmargin=*,noitemsep,topsep=0pt]
    \item \texttt{EdgeRemoving}: drops edges from a node's neighborhood to simulate minor, non-critical logical bypasses or ``don't care'' conditions. This enables the model to learn that a latch's core function should be robust to such local rewiring.
    \item \texttt{FeatureMasking}: zeros out feature dimensions to force the model to learn from a more holistic set of features, reducing its reliance on any single characteristic and promoting a focus on the broader topological context.
\end{itemize}

For the training of this GraphCL model, we make use of 390 circuit graphs with latch count below one thousand.
This contrastive learning task forces the model to learn what is essential about a latch's role, producing a dense vector embedding that encapsulates its rich, functionally-relevant topological context.\looseness=-1

\subsubsection{Augmentation of Latch Embeddings.}

In addition to the embeddings produced by GraphCL, which represents static information in a circuit graph, we also incorporate dynamic functional behaviors of latches.
Specifically, we augment the GraphCL embeddings with a dynamic functional feature: the \textit{signal flip rate}, $r_{\text{flip}}$. We perform a fast bounded logic simulation and record the normalized frequency at which each latch toggles its state over $T$ random cycles:
\begin{equation}
    r_{\text{flip}}(l) = \frac{1}{T} \sum_{t=1}^{T} \mathds{1}[s_t(l) \neq s_{t-1}(l)]
\end{equation}
where $s_t(l)$ is the state of latch $l$ at cycle $t$. This value, serving as a proxy for the latch's activity, is concatenated with its learned structural embedding. The representation for each latch is a vector $v_l = [e_l, r_{\text{flip}}(l)] \in \mathbb{R}^{d+1}$ that captures both its static structural role and dynamic functional activity.


\subsection{Clause Prediction and Integration}
With a pre-computed embedding for every latch, the expensive per-clause graph-based analysis of prior works is no longer needed. This enables an extremely fast and lightweight inference stage.\looseness=-1

\subsubsection{ML-aided Inductive Generalization.}

Leveraging pre-computed embeddings, the final challenge is to design a machine learning model that can effectively score each literal based on its role in the circuit. A crucial insight is that a clause is an \textit{unordered set of literals}.
\legend addresses this permutation-invariance using a lightweight DeepSets‑style model~\cite{deepsets} (a permutation‑invariant network over sets that aggregates element features), which first aggregates the embeddings of the literals into a global, order‑agnostic representation, and then uses this aggregated representation to score each literal. Let \(C = \{\ell_1, \dots, \ell_m\}\) be a minimized CTI, where some literals can be  dropped to form a generalized clause. The machine learning model now will try to predict the literals that can be dropped.
Let \(v_i \in \mathbb{R}^d\) denote the pre‑computed augmented embedding of the latch in literal \(\ell_i\) (where \(d\) is the embedding dimension). The model operates in two steps:
\vspace{-5pt}
\[
\underbrace{%
g \;=\;\rho\!\left(\textstyle\sum_{i=1}^{m}\phi(v_i)\right)}_{\text{set\,–\,level aggregation}}
\quad;\qquad
\underbrace{%
s_i \;=\; \psi\!\bigl([\,v_i \,\|\, g\,]\bigr)\quad\text{for }i=1\dots m}_{\text{local\,\&\,global scoring}}
\]
\begin{itemize}[leftmargin=*,noitemsep,topsep=0pt]
\item \textbf{Set aggregator:}  We use a shared multi-layer perceptron (MLP) $\phi:\mathbb{R}^{d}\!\to\!\mathbb{R}^{h}$ to transform each literal's embedding independently; the sum $\sum_i\phi(v_i)$ (a symmetric operation) guarantees permutation invariance. A second MLP $\rho:\mathbb{R}^{h}\!\to\!\mathbb{R}^{h}$ yields the global clause representation $g$, where $h$ denotes the hidden/global dimension; hence $g\in\mathbb{R}^{h}$.
\item \textbf{Local–global scoring:} For each literal, we concatenate its embedding \(v_i\) with the global vector \(g\), forming the input \([v_i \| g] \in \mathbb{R}^{d+h}\), which is passed through another MLP \(\psi: \mathbb{R}^{d+h} \to [0, 1]\). The outputs \(s_i\) are treated as  scores of the literals that will decide if a literal shall be kept or dropped from the CTI to form a generalized clause. In addition, $d$ and $h$ are fixed across experiments.
\end{itemize}


In the training phase, the above MLPs are trained using the clauses in the known inductive invariants in the training set. Namely, for an CTI $s$, there must exist (at least) one clause $c$ in the inductive invariant such that $c \Rightarrow \neg s$. When there are multiple clauses, we always use the one discovered at the latest frame. We then assign binary labels to literals in $C$: keep if the literal appears in $c$, and drop otherwise. The predictor is trained to approximate these keep/drop decisions.\looseness=-1

Given a minimized CTI $C=\{\ell_1,\dots,\ell_m\}$, the predictor learns a permutation‑invariant, clause‑conditioned scoring function $s(\ell_i\mid C)\in[0,1]$ that estimates the necessity of retaining literal $\ell_i$ so that, after dropping low‑scoring literals, the resulting clause remains valid under the clause sanity checks introduced in \Cref{sec:method:cslsideloading}. The set aggregator conditions each literal on the rest of $C$, capturing co‑occurrence and redundancy among literals. Each literal is represented by a latch embedding that summarizes its structural context together with a dynamic flip‑rate feature as described above, providing the evidence used by the scoring function.


Since the embeddings of literals are processed independently, the entire process is permutation-invariant by design.
Moreover, the above inference-time machine learning pipeline contains \emph{no} graph convolutions or topological message passing, so inference remains
orders‑of‑magnitude faster than previous GNN‑based approaches while still respecting the
set nature of IC3 clauses.

\input{tables/overall_comparison_with_deepic3}

\subsubsection{Clause side-loading.}\label{sec:method:cslsideloading}

Finally, each candidate clause is instantiated by retaining the literals with scores $\ge \theta$, where $\theta$ denotes the literal‑selection threshold; during inference, if no candidate survives at the current $\theta$, we relax it via a simple adaptive decay. Clauses for which all literals fall below the threshold are discarded, yielding the selected set $\mathcal{C}_{\text{selected}}$.
However, these clauses may not satisfied the requirements of clauses in the PDR frames. 
To guarantee the logical soundness of the provided lemmas, we subject each candidate in $\mathcal{C}_{\text{selected}}$ to a rigorous formal \textit{sanity check}. A clause $C$ is only accepted if it satisfies the following two fundamental properties of a clause in frame $F_1$:

\begin{itemize}[leftmargin=*,noitemsep,topsep=0pt]
    \item \textbf{Initiation:} The clause must hold in all initial states (${I} \models C$), which is verified by ensuring ${I} \land \neg C$ is unsatisfiable.
    \item \textbf{Consistency with the 1$^{st}$ transition:} The clause must hold for all states that are reachable following one state transition  $T$ from the initial states, namely $I \land T \models C'$. This is verified by checking if $I \land T \land \neg C'$ is unsatisfiable. Here $C'$ denotes the clause $C$ with all its state variables advanced to the next time-step.
\end{itemize}
Only clauses that pass this sanity check form the final set $\mathcal{C}^*$:
\begin{equation}
\mathcal{C}^* = \{ C \in \mathcal{C}_{\text{selected}} \mid ({I} \models C) \land (I \land T \models C') \}
\end{equation}
We refer to clauses that satisfy both properties as \emph{sanity‑checked} clauses. 
These are the minimal requirements for a clause to hold at the frame $F_1$. These sanity‑checked clauses are then side-loaded into the IC3/PDR solvers at initialization~\cite{Hu21}. By providing the solver with these clauses, we give it a powerful head-start, effectively guiding its search toward a quicker proof or refutation and drastically accelerating its convergence.\looseness=-1

%% file: tables/overall_comparison_with_deepic3.tex
\newcommand{\best}[1]{\cellcolor{blue!8}\textbf{#1}}

\begin{table*}[h]
\setlength{\tabcolsep}{8pt}
\small
\centering
\begin{threeparttable}
\caption{Wall-clock time comparison of IC3/PDR engines under different lemma-generation strategies\dag}
\begin{tabular}{@{\hspace{\fboxsep}}lcccccc@{\hspace{\fboxsep}}}
\toprule
\textbf{Method} & \textbf{\# Solved/Total} & \textbf{\# SAFE Solved} & \textbf{\# UNSAFE Solved} & \textbf{Total-PAR2 (h)} & \textbf{Avg-PAR2 (s)} & \textbf{PAR2 Speedup} \\ \midrule\midrule
IC3ref-portfolio\ddag & 166/200 & 128 & 38 & 176.32 & 3173.69 & 1.00 \\
IC3ref with \texttt{DeepIC3}~\cite{Hu21} & 167/200 & 129 & 38 & 173.88 & 3129.87 & 1.01 \\
IC3ref with \texttt{CTP}~\cite{predictinglemmasgeneralizationic3} & 166/200 & 130 & 36 & 172.51 & 3105.14 & 1.02 \\
IC3ref with \legend* & \best{181}/200 & \best{138} & \best{43} & \best{112.69} & \best{2028.50} & \best{1.56} \\ \midrule
ABC         & 160/200 & 127 & 33 &192.15  & 3458.65 & 1.00 \\
ABC with \texttt{DeepIC3}~\cite{Hu21} & 160/200 & 127 & 33 & 191.63 & 3449.39 & 1.00 \\
ABC with \legend*  & \best{182}/200 & \best{134} & \best{48} & \best{108.23} & \best{1948.14} & \best{1.78} \\ \bottomrule
\end{tabular}
\begin{tablenotes}[flushright]
  \small
  \item[\dag] Timeout for each case is set to 7200 (s). PAR2 counts each timeout as twice the time limit; totals/averages are computed under this rule.
  \item[\ddag] IC3ref-portfolio contains two modes with basic generalization enabled or disabled, where the better result is recorded.
  \item[*] \legend timings include side-loading overhead, i.e., time spent producing clauses and performing sanity checks in addition to solver runtime.
  \item[\S] ABC with CTP results are omitted because the authors have not released an ABC integration of CTP.
\end{tablenotes}
\label{tbl:overall_comparison}
\end{threeparttable}
\vspace{-10pt}
\end{table*}

%% file: sections/experiment.tex
\section{Experiments}\label{sec:experiment}
\subsection{Experiment Setting}\label{sec:exp_setting}

\subsubsection{The Platform of Experiments.}
All experiments are performed on a Ubuntu 20.04.4 LTS server with a single NVIDIA GeForce RTX~3090 GPU and dual Intel Xeon Platinum 8375C processors (256\,GB of memory). Our framework \legend is implemented in C++ and Python.
We use two widely‑used PDR solvers in the experiments, namely the PDR engine implemented in ABC~\cite{abc} and a standalone PDR reference implementation IC3ref~\cite{ic3ref_repo}.
For clause side-loading, we implement a simple interface in these two tools to allow passing predicted clauses into the frame $F_1$ at initialization.


\subsubsection{Benchmark Selection.}
To construct a comprehensive and challenging evaluation suite, we begin with all publicly available single-property safety benchmarks from the Hardware Model Checking Competition (HWMCC) archives~\cite{hwmcc} spanning from the year 2008 to 2024. From this initial large pool of model checking problems, we curate the training set and test set following the following criteria:\looseness=-1

%

\begin{itemize}
\item \textbf{Training set with clause‑level labels.} For training set,  we first filter out those known unsafe cases as they have no inductive invariants, as well as those trivially safe cases (when the property itself is inductive) which require no additional inductive clauses. 
For the remaining problems, to make training cost moderate, we set a limit to the latch count to be below 20000 and sample around 250 AIGs from the pool. We then run an oracle PDR engine with 7200-second timeout. For those proved safe cases, we extract the clauses in the inductive invariant and sample the CTIs as described earlier. This yields approximately $1.79\times 10^6$ training labels.
\item \textbf{Test suite.} For evaluation, we cast a much wider suite. Any circuit that can be embedded using GraphCL on our 24GB-RAM GPU (using FP16 format) is eligible, regardless of latch count. This allows us to probe \legend's ability to scale beyond its training regime. We remove those already used in the training set, and then from the remaining pool,  randomly sample 200 instances to form the final test suite. In approximate terms, the suite follows a roughly \(2\!:\!4\!:\!2\!:\!2\) split across four runtime bands—within 60~s, 60 to 1000~s, 1000 to 7200~s, and above 7200~s, providing a broad yet balanced mix for assessing both speed‑ups and scalability of \legend. The test suite is checked to be strictly disjoint from the training set.\looseness=-1
\end{itemize}

\subsection{Results and Analysis}
\textbf{\Cref{tbl:overall_comparison}} illustrates the end‑to‑end result of side-loading lemmas generated by \legend  into the two PDR engines, IC3ref~\cite{ic3ref_repo} and ABC~\cite{abc}, and it also compares with two modern lemma‑generation techniques (DeepIC3~\cite{Hu21} and IC3-CTP~\cite{predictinglemmasgeneralizationic3}) for reference. The time limit is 7200 seconds (wall-clock time) and all \legend timings include clause production and sanity checks.

The first observation of \legend side-loading is a markedly higher proof rate. IC3ref with \legend now concludes on 181 of 200 instances, \textbf{15} more than the vanilla IC3ref; ABC with \legend solves 182 of 200, a gain of \textbf{22} over its baseline. Crucially, the additional successes are split across both SAFE and UNSAFE categories. Successful SAFE proofs rise from 128 to 138 for IC3ref and from 127 to 134 for ABC, while UNSAFE findings increase by \textbf{5} and \textbf{15} cases respectively. \emph{By contrast, DeepIC3 and IC3-CTP improve the proof rate only marginally (at most +1 instance on IC3ref and 0 on ABC) and yield near‑neutral PAR‑2 improvements (about 1.00 to 1.02$\times$), with no consistent gains on UNSAFE cases}.

Regarding the PAR‑2 metric, which penalizes every timeout case by counting twice the time limit, the benefits become even clearer. 
For IC3ref the cumulative PAR‑2 time drops from 176.32~h to 112.69~h, yielding a \textbf{1.56$\times$} speed‑up.
The effect is similarly strong for ABC: total PAR‑2 time shrinks from 192.15~h to 108.23~h, yielding a \textbf{1.78$\times$} average speed‑up.
\begin{figure}[b!]
    \centering
    \includegraphics[width=1\columnwidth]{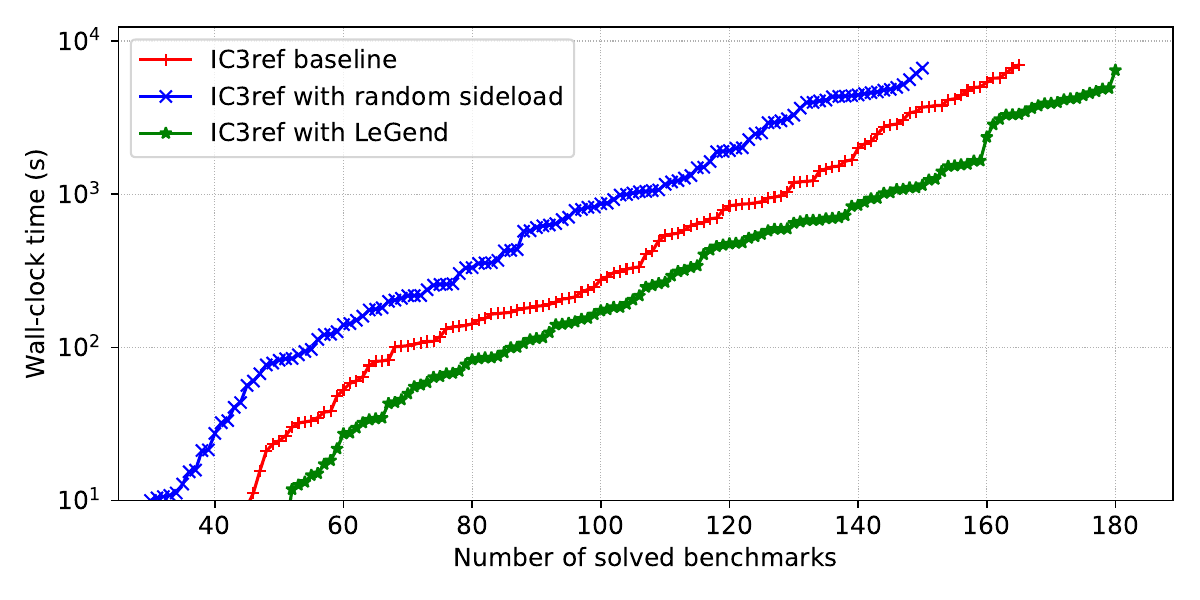}
    \caption{Ablation study of \legend side-load lemma quality
    }
    \label{fig:ablation}
\end{figure}
From this experiment, we would like to draw the following two conclusions. 
First, the effectiveness of predicted lemmas is not limited to SAFE cases, where inductive invariants are expected. They can also accelerate the discovery of genuine counter‑example traces when the property is invalid. Second, the fact that two IC3 engines with different clause‑management heuristics improve by similar margins suggests that the lemmas capture design‑level regularities rather than overfitting to a specific solver.

\subsection{Ablation of Side-Loaded Lemma Quality}

To further evaluate the effectiveness of \legend's predicted lemmas, we perform an  ablation study with three configurations of IC3ref on the same 200‑circuit benchmark suite.

As illustrated by \textbf{\Cref{fig:ablation}}, the red curve denotes the vanilla IC3ref solver (the baseline); the green curve augments it with \legend's lemmas; the blue curve is a randomized control, where we obtain clauses by randomly retaining or discarding literals in the minimized CTIs, and then side-load only those that pass the same sanity checks to the PDR solver.

The outcome is unambiguous. Random side-loading not only fails to accelerate verification, it actively hurts: the blue curve lags behind the baseline across almost the entire time axis and terminates with fewer solved instances, confirming that indiscriminate lemma side-loading is not beneficial and a good prediction of clauses is needed.
In contrast, \legend's curve stays to the right of both competitors, solving more benchmarks at every time budget and ultimately adding \textbf{15} extra proofs over the baseline. 
This ablation study illustrates that the speed-up we observe in the experiment is not the direct result of clause side-loading and it confirms the effectiveness of clause prediction of our machine learning model.


%% file: sections/conclusion.tex
\section{Conclusion and Future Work}
\label{sec:conclusion}

This paper addresses the critical scalability bottleneck of prior machine-learning-guided PDR algorithms through \legend, a novel framework equipped with an efficient one-time global representation learning approach.
\legend delivers substantial performance acceleration to the state-of-the-art IC3/PDR solvers with negligible inference overhead, as demonstrated by our experiments. Compared with the marginal gains typically achieved by traditional algorithmic tuning, machine-learning-guided IC3/PDR can achieve considerable speed-ups, highlighting a powerful and promising direction for future model-checking advancements. Our current implementation utilizes only a single GPU. Scaling the framework to multi-GPU training and inference in the future is expected to further amplify these performance gains.\looseness=-1